# Teaching Reform and Exploration on Object-Oriented Programming

Guowu Yuan[1], Bing Kong[1], Haiyan Ding[1], Jixian Zhang[1], Yang Zhao[2]
*1 School of Information Science & Engineering, Yunnan University, Kunming, China*
*2 Department of Animation, Yunnan Normal University, Kunming, China*

*Abstract*—**The problems in our teaching on object-oriented programming are analyzed, and the basic ideas, causes and methods of the reform are discussed on the curriculum, theoretical teaching and practical classes. Our practice shows that these reforms can improve students' understanding of object-oriented to enhance students' practical ability and innovative ability.**

## I. INTRODUCTION

Programming is the most important and essential ability for the undergraduate majored in computer science and technology. With the development of programming language, structured programming has been replaced with object-oriented programming, and object-oriented programming becomes main idea of programming[1]. It has become an urgent problem to improve the students' ability of object-oriented design and programming.

## II. PROBLEMS IN THE TEACHING

Yunnan University has offered object-oriented programming in computer science and technology from 1998. This course is offered as a professional elective course in the 6th semester with 3 credits (54 class hours for theory, 28 class hours for experiment).

According to our teaching experiences in many years, there are primary problems existed in the course as follows:

(1) Students have been familiar with the structured programming which is represented by C language, and therefore they are difficult to alter their ways for object-oriented programming. They have learned C language in the first semester and data structure in the second semester. With a large number of programming with C language, they have been familiar with the process of structured programming. When they use object-oriented programming language, they are difficult to accept the idea and method of object-oriented.

(2) In the theory courses, the relevance between the examples is low. In the experimental courses, verification experiments are too much and simple, and the relevance and continuity between previous and posterior experiments is low, which badly affects understanding object-oriented programming on the whole. All experimental exercises are some very small "toy" programs, and thus students cannot be trained using comprehensive experiments.

(3) In the China Rank Certificate Test of Computer Software, our undergraduates cannot have good grades in the object-oriented programming. The China Rank Certificate Test of Computer Software is a professional certification examination of computer, and it can reflect the cultivation of students' comprehensive ability. The certification is an important index for employers to pick students. With our experience for marking the certification papers of Yunnan Province, we found that oriented-object programming in the software designer is not good, the majority of students have a score lower than 1/3 of the total score. Therefore, it is very necessary to strengthen students to train object-oriented programming.

In summary, it is very important to reform the course teaching of the object-oriented programming in order to improve students' object-oriented programming ability and to enhance the passing rate of the China Rank Certificate Test of Computer Software.

### III. ASURES AND METHODS OF TEACHING REFORM

#### A. Curriculum adjustment

In our revised teaching plan, object-oriented programming is offered in the third semester ahead of the old plan. The course uses Java as an example to explain and experiment, with 3 credits (36 class hours for theory, 36 class hours for experiment). The main reasons for the curriculum adjustment are as follows:

(1) The object-oriented programming languages, which are represented by Java, are widely used. Object-oriented programming is an indispensable skill for the students majored in computer science, and then it is taught early to enable students to have more time to study;

(2) Java language can be used as the experimental tool in other professional courses, such as in certain knowledge point of each course in the list in Table I, these knowledge can be convenient to experiment using Java;

(3) In the arrangement of professional course, at least one course, which is related with programming, is offered in each semester. programming courses are continued in the all four years of undergraduate. The courses with programming experiment are shown in the table II. As can be seen, the object-oriented programming is offered in the third semester and will be beneficial to the subsequent courses.

TABLE I. THE APPLICATION OF JAVA IN EACH COURSE EXPERIMENT

| Course | Experiment content | Java API |
|---|---|---|
| Data structure | Linear list | ArrayList、LinkedList |
| | Stack | Stack |
| | Queue | Queue |
| | Sort | Collection.sort()、Arrays.sort() |
| | Search | Collection.binarySearch()、Arrays.binarySearch () |
| | Hash table | HashSet、HashMap |
| | Sort tree | TreeSet、TreeMap |
| Operating system | Process operation | Process、ProcessBuilder |
| | Thread operation | Thread, Runnable, synchronized, wait（）, notify, notifyAll（）, etc. |
| | I/O stream | InputStream, OutputStream, etc. |
| | Priority | MIN_PRIORITY、NORM_PRIORITY、MAX_PRIORITY in Thread |
| Database | Interface design of management | AWT, Swing, etc. |

|  | information system |  |
|---|---|---|
|  | Database operation of management information system | JDBC |
| Computer network | TCP | Socket |
|  | UDP | DatagramPacket |
|  | Data broadcast | MulticastSocket |

TABLE II. THE PROGRAMMING COURSE IN EACH SEMESTER

| Semester | Course |
|---|---|
| 1 | Programming foundation (C language teaching) |
| 2 | Data structure |
| 3 | Object-oriented programming (Java language teaching) |
| 4 | Computer graphics, computer network, assembly language programming |
| 5 | Operation system, algorithm design and analysis, distributed software development technology |
| 6 | Compiling technology, database design and Application |
| 7 | Software engineering practice, multi-core program design |
| 8 | Graduation design |

## B. The reform of teaching theory

In the teaching of object-oriented programming using Java, the traditional method is to explain the history and the basic syntax of Java language, and then to explain theoretical knowledge of object-oriented, class, object, encapsulation, inheritance and polymorphism, which is the conventional books' arrangement order.

But for many schools, the C language is the first programming language for student. They have been accustomed to procedure programming idea. Because teaching examples are very simple, these examples can also be very easy to code using C language. The code using Java is more complex and longer, students will consider the object-oriented programming makes simple problems complicated.

In fact, students have already had the idea of the structured programming, and this course should solve how to change the students' thinking to the object-oriented. It can achieve better result to adjust the order of teaching content. We take the following teaching order:

(1) The difference of basic syntax between Java and C language. Because the basic syntax of Java and C are mostly similar and students have the C language foundation, students only need to study the differences. The difference of basic syntax between Java and C language are mainly the following:
- The basic data types: (a) The char type: An character is 1 bytes in C language using the ASCII code, but is 2 bytes in Java using Unicode code; (b) the Boolean type: in C language, zero means false, and non-zero means true; but Java provides the basic data types of Boolean with only two kind of values: false and true.
- Array type: In the C language, arrays can be used after definition, but in Java, arrays must be defined and created before they can be used.

TABLE III. THE DIFFERENCE OF BASIC SYNTAX BETWEEN JAVA AND C LANGUAGE

| Programing language | C | Java |
|---|---|---|
| byte | ASCII(8 bit) | Unicode(16 bit) |
| Boolean | zero means false, and non-zero means true | false or true |
| array | Do not need to create before use | Need to be created before use |

(2) Explain several Java system class, let students know the benefits of object-oriented programming. Several Java system class are chosen to make students familiar, let students know the benefits of object-oriented programming by the implementation of some small programs [2]. For example:
- String: String operations are the most commonly used in programming. Using C language, a lot of code needs to be written. But using Java, a variety of string operations can be done by calling the system class String using the format "object.method (parameter)". Students do not need to understand how these methods to achieve, just know how to call and parameters needed.
- Frame or JFrame: Now, most of programs have visual user interfaces. After learning C language, because students have no way to achieve a visual user interface, students have been very curious about how to achieve a visual user interface. After the basic methods of Frame or JFrame in Java are explained to students, the students can use "JFrame myWindow = new JFrame();" to create a display window, and then they can use "myWindow.setTitle (" My Window");" to set the window title, use "myWindow.setSize (800,600);" to set the size, also can change the background of the window, add a label to the window etc.. These operations only need a few statements to create a visual window, do not need to understand how to achieve Frame or JFrame. Students will feel that object-oriented programming is very simple and convenient.

(3) Object-oriented knowledge. Students experience the convenience and simplicity of a few Java system class, and they are interested in learning how to define the classes they need. When we explain it, it is important to choose a good example.
- The most common example is to calculate graphics area. For example, a class Circle can be defined, it has public member variable radius and member method getArea(). The variable radius represents the radius of the circle, and the method getArea() calculates the area of the circle. Then, an object of class Circle, named by c, is defined, the object c can directly assigned by "c.radius=10;". But by this way, the member variable radius can be given by a negative number, which does not obviously meet the definition of a circle. At this time, students can be guided to set radius to private types. Because to directly access private variable is not allowed, the public method setRadius() is added to set the radius. In the method setRadius(), if the given radius is negative, the radius is set to 0. After learning the exception handling, it can be modified to throw an exception when the radius is less than 0. This can guarantee the radius to get a reasonable value, can ensure the correctness and legitimacy of the program. Through this example, the concept of encapsulation can be introduced.
- After that, since it is not convenient to use setRadius() to set radius after each circle is defined, the concept of the construction method can be derived. With the construction methods, it is convenient to initialize objects.
- Subsequently, in order to record the number of circle's objects generated, static variables are introduced.
- In the interpretation of inheritance, the class Cylinder can be introduced. Because the bottom surface of a cylinder is a circle, the cylinder's height is added on the basis of the class Circle. When the cylinder volume is calculate, the method getArea() in the class Circle can be used. Thus, the concept of inheritance can be introduced. Of course, it is the best to use the combination of classes to achieve this example. It can be compared with the inheritance of class.

- In explaining the polymorphism, other graphs are introduced, such as triangles, rectangles, and their methods calculated their areas are unified to the method name of getArea(). Then, abstract class Graphics is defined as their superclass, and abstract method getArea() is defined in the abstract class Graphics.
- In the subclass, the method for initializing object is also needed, and then the construction method of the subclass can be explain using the keywords super.
- In this way, we can use the graph area calculation to lead to many important concepts, and through this example, we can see the link between these main concepts. When we explain graphic user interface, we can ask students to design a graphic interface program to calculate all kinds of figures' area, which is based on the above.

*C.  The reform of the practice teaching*

In making reform in theoretical class, the reform of the practice is more important. Object-oriented programming has lots of knowledge points and wide applications. Therefore, multi-level practice teaching system can be set up as follows.

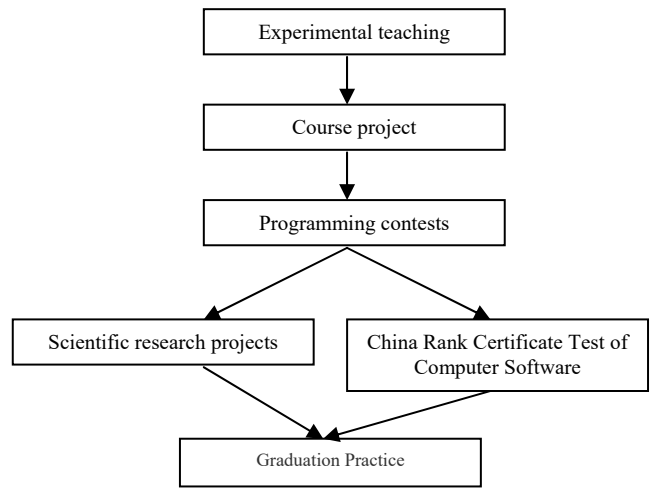

FIGURE I. PRACTICE TEACHING PROCESS

- The experiments in the computer room. The experiments mainly verify each knowledge point. Designing the content of the experiments, we try to make it possible to have an association between the front and the back of the experiment. The back experiments are to perfect the front experiments, and eventually students can get a small application system. In the experiments, we can select programming questions from the China Rank Certificate Test of Computer Software, and all students must complete the questions. Students will advance into the actual combat of the China Rank Certificate Test of Computer Software.
- Course project. We design some slightly larger and difficult topics which have the comprehensive use of knowledge. A team with 3~5 students completes a topic and finally explains it. According to the completion of each team, we make a comprehensive evaluation to each team.
- Programming contests, scientific research projects to undergraduate and the China Rank Certificate Test of Computer Software. After learning the course, students are required to take part in the annual test of programmers or software designers in the China Rank Certificate Test of Computer Software. We give the relevant training to the students with strong learning ability, and encourage them to participate in the ACM International College Programming Contest, Baidu star

- programming contest, TopCoder programming contest, Oracle ThinkQuest programming contest, etc. The students with creative thinking are encouraged to join provincial or national undergraduate research project.
- Graduation Practice. Before graduation, the students do graduation Practice in the software company, and they can enhance the ability of the comprehensive programming

After the above practices, the students will have a great improvement in the object-oriented programming, and will have better employment prospects.

## IV. CONCLUSIONS

Object-oriented programming is a basic course for computer specialty, and it has more help to other courses' study and experiment. Object-oriented programming is the mainstream of programming, and programming capacity is the maximum reflect the ability of students majored in computer, so the course is very important.

This paper analyzes the existed problems of the course teaching and proposes the relevant reform measures, including the reform of curriculum, theory course and practice course. These reforms can give some inspiration to relevant teachers. We will carry out teaching reforms in natural language processing [3-4], image processing [5], AI-based education [6], and deep learning [7].

### ACKNOWLEDGMENT

This work is supported by the teaching reform research project of Yunnan University(No. 2016024), the Application and Foundation Project of Yunnan Province (No. 2015FB115), and the Young Teachers Training Program of Yunnan University.

### REFERENCES

[1] Zhi-guo Ding, Jie Qian, "Teaching reform on object-oriented programming course", Computer Education, 2011(9):9-12.

[2] Jian Shu, Wen-yong Wen, "Teaching practice and thinking of Java programming," Computer Education, 2008(24):147-149+62.

[3] Ngo C W , Jiang Y G , Wei X Y ,et al.Beyond Semantic Search: What You Observe May Not Be What You Think[C]//TRECVID 2008 workshop participants notebook papers, Gaithersburg, MD, USA, November 2008.DBLP, 2008.

[4] Chong-wah Ngo Shi-ai Zhu Hung-khoon Tan Wan-lei Zhao Xiao-yong Wei.VIREO at TRECVID 2010: Semantic Indexing, Known-Item Search, and Content-Based Copy Detection[C]//Trecvid Workshop Participants Notebook Papers.DBLP, 2010.

[5] Ngo C W , Pan Z , Wei X ,et al.Motion Driven Approaches to Shot Boundary Detection, Low-Level Feature Extraction and BBC Rushes Characterization at TRECVID 2005[J].trecvid workshop, 2005.

[6] Wei X Y , Yang Z Q .Mining in-class social networks for large-scale pedagogical analysis[C]//Proceedings of the 20th ACM international conference on Multimedia.ACM, 2012.DOI:10.1145/2393347.2393436.

[7] Wei X Y , Yang Z Q , Zhang X L ,et al.Deep Collocative Learning for Immunofixation Electrophoresis Image Analysis[J].IEEE transactions on medical imaging, 2021, 40(7):1898-1910.DOI:10.1109/TMI.2021.3068404